\documentclass[twocolumn]{aastex61}
\usepackage{natbib}
\usepackage{amsmath}
\newcommand{\beq}{\begin{equation}}
\newcommand{\eeq}{\end{equation}}
\newcommand{\bal}{\begin{align}}
\newcommand{\eal}{\end{align}}

\renewcommand{\vec}[1]{\textbf{#1}}

\received{2018 January 10}
\revised{2018 July 2}
\accepted{2018 July 4}

\shorttitle{Sub-grid model for dust growth}
\shortauthors{Tamfal et al.}

\begin{document}

\title{A sub-grid model for the growth of dust particles\\
 in hydrodynamical simulations of protoplanetary disks}

\correspondingauthor{Tomas Tamfal}
\email{tomas.tamfal@uzh.ch}

\author[0000-0003-1773-9349]{Tomas Tamfal}
\affil{Center for Theoretical Astrophysics and Cosmology, Institute for Computational Science, University of Zurich, Winterthurerstrasse 190, CH-8057 Z\"urich, Switzerland}

\author[0000-0002-9128-0305]{Joanna Dr\c{a}\.{z}kowska}
\affiliation{Center for Theoretical Astrophysics and Cosmology, Institute for Computational Science, University of Zurich, Winterthurerstrasse 190, CH-8057 Z\"urich, Switzerland}

\author[0000-0002-7078-2074]{Lucio Mayer}
\affiliation{Center for Theoretical Astrophysics and Cosmology, Institute for Computational Science, University of Zurich, Winterthurerstrasse 190, CH-8057 Z\"urich, Switzerland}

\author[0000-0001-8998-72674]{Clement Surville}
\affiliation{Center for Theoretical Astrophysics and Cosmology, Institute for Computational Science, University of Zurich, Winterthurerstrasse 190, CH-8057 Z\"urich, Switzerland}

\begin{abstract}

We present the first 2D hydrodynamical finite volume simulations in which dust is fully coupled with the gas, including its back-reaction onto it, and at the same time the dust size is evolving according to coagulation and fragmentation based on a sub-grid model. 
The aim of this analysis is to present the differences occurring when dust evolution is included relative to simulations with fixed dust size, with and without an embedded Jupiter-mass planet that triggers gap formation. 
We use the two-fluid polar Godunov-type code RoSSBi developed by \cite{Surville2016} combined with a new local sub-grid method for dust evolution based on the model by \cite{Birnstiel2012}. 
We find striking differences between simulations with variable and fixed dust sizes.
The timescales for dust depletion differ significantly and yield a completely different evolution of the dust surface density.
In general sharp features such as pile-ups of dust in the inner disk and near gap edges, when a massive planet is present, become much weaker.
This has important implications on the interpretation of observed substructure in disks, suggesting that the presence of a massive planet does not necessarily cause sharp gaps and rings in the dust component.
Also, particles with different dust sizes show a different distribution, pointing to the importance of multi-wavelength synthetic observations in order to compare with observations by ALMA and other instruments.
We also find that simulations adopting fixed intermediate particle sizes, in the range $10^{-2} - 10^{-1}$ cm, best approximate the surface density evolution seen in simulations with
dust evolution.

\end{abstract}


\section{Introduction}\label{s:intro}
Over the last decade the study of protoplanetary disks has become an increasingly important research topic in astrophysics owing mostly
to the tremendous progress made by multi-wavelength high-resolution observations that can finally reveal their internal
structure.
In particular, $\sim$ AU-resolution interferometric observations of millimeter sized grains, which are partially decoupled from the gas, in systems such as HL Tau and TW Hydra, 
provided by ALMA (see \citealt{Partnership2015} and \citealt{Andrews2016}) are allowing us to investigate in detail the dust distribution 
within such disks. On the other hand, data from the SPHERE telescope bring information about the small particles which are well coupled with the gas, e.g. TW Hydra from \cite{Menu2014} or \cite{Boekel2017}.
These observations display that disks are not homogeneous in structure, rather they exhibit axisymmetric rings and gaps, as well as,
in other cases, spiral structure and other non-axisymmetric structures (\citealt{Benisty2015}, \citealt{Dong2017}).
Additionally, they suggest that the radial distribution of dust sizes estimated by the spectral index is correlated with the change of brightness (\citealt{Partnership2015}).
These features have been found in recent simulations (\citealt{Flock2015}, \citealt{Dong2017}).
In order to explain these observations, various numerical modeling approaches have been attempted, and different possible 
scenarios to reproduce the observations have been suggested, such as planets generating the gaps and neighboring pressure
bumps, disk instabilities, planet-triggered spiral density waves, or the variation of dust properties at the snow lines
(\citealt{Fouchet2010}, \citealt{Flock2015}, \citealt{Dong2017}).
A powerful numerical tool are hydrodynamical simulations, which normally follow the gas and dust components but
assume a fixed dust size (\citealt{Dipierro2015}) and often neglect the back-reaction of dust onto gas (\citealt{Picogna2015}).
In the literature there are two main approaches to solve the hydrodynamical equations, one is a grid-based approach, e.g. \cite{Stone2008}, and the other is a particle-based approach (hereafter SPH), e.g. \cite{Gonzalez2012}.
An additional possibility to explain the observations is a one dimensional analysis which can be used to investigate dust growth within protoplanetary disks 
(e.g. \citealt{Zhang2015}, \citealt{Okuzumi2016}).
The main focus of this work will be the analysis of protoplanetary disks containing fully coupled gas and a dust fluids with an evolving dust size for now
limited to a two dimensional configuration. We will highlight the differences, for various setups, with and without embedded planets,
arising between models with fixed dust size and models with an evolving dust size. We will also compare our results for fixed dust size
with previously published results, for example those recently obtained with the ATHENA code by \cite{Zhu2014}.

It may seem unnecessary to focus on dust coagulation, since a disk is composed of approximately 99$\%$ of gas and roughly 1$\%$ of dust. 
However, despite its insubstantial fraction, dust does influence the disk evolution. 
Dust particles play a crucial role in the planetesimal growth; hence simulating solid particles within the disk is important for understanding planet formation. 
In order to simulate the dust coagulation we used the finite volume Godunov-type code RoSSBi described in \cite{Surville2016}.
Similar studies have been performed with a dust coagulation scheme incorporated into an SPH simulation, for example in \cite{Gonzalez2015},
but this is the first study of such type using a finite volume code. 
In contrast to these simulations the RoSSBi code is a Godunov-type finite volume method and hence uses a different strategy to simulate such problems.
The simulation of dust coagulation ab initio and self-consistently, within a two dimensional framework, 
is not possible even with state-of-the-art parallel computing architectures since it would require adding the dust size 
dimension and solving the Smoluchowski equation on its own grid of dust sizes nested into individual cells of the hydrodynamical grid.
For now, this is computationally feasible only in a one-dimensional (radial) approach, 
and using implicit integration schemes (\citealt{Brauer2008} and \citealt{Birnstiel2010}) that are different from integration schemes adopted for solving dust advection.

Therefore in this paper we model dust fully coupled with the hydrodynamics but in order to follow its evolution 
we employ a relatively simple sub-grid method based on the two population algorithm proposed by \cite{Birnstiel2012}. This
approach lowers the computational cost, thus making implementation of dust growth in an advanced 2D or 3D simulation possible. 

\section{Methods}
\subsection{Two-fluid simulation technique}

In this work we employ the finite volume code RoSSBi (\citealt{Surville2015, Surville2018}), 
which uses the fluid approximation to treat both the gas and the dust component of the disk,
solving the relevant equations in two dimensions and in cylindrical coordinates.
The evolution of the dust and gas surface density ($\Sigma_{d} $, $\Sigma_{g} $) is 
described by the inviscid Euler equations in a cylindrical coordinate system. For the gas surface density one obtains
\beq
\frac{\partial}{\partial t} \Sigma_{g} + v_{g,r} \frac{\partial}{\partial r} \Sigma_{g} + \frac{v_{g,\phi}}{ r} \frac{\partial}{\partial \phi} \Sigma_{g} + \Sigma_{g} \nabla \cdot  \vec{v}_{g} = 0,
\eeq
with $v_{g,r}$, $v_{g,\phi}$, and $\vec{v}_{g}$ denoting the radial, azimuthal and total gas velocity, respectively.
The evolution of the dust surface density reads
\beq
\frac{\partial}{\partial t} \Sigma_{d} + v_{d,r} \frac{\partial}{\partial r} \Sigma_{d} + \frac{v_{d,\phi}}{ r} \frac{\partial}{\partial \phi} \Sigma_{d} + \Sigma_{d} \nabla \cdot  \vec{v}_{d} = 0,
\eeq
with $v_{d,r}$, $v_{d,\phi}$, and $\vec{v}_{d}$ denoting the radial, azimuthal and total dust velocity, respectively. 

In absence of additional force, the evolution of the gas velocity field follows
\beq\label{eq:gasevo}
\frac{\partial \vec{v}}{ \partial t} + ( \vec{v} \cdot \nabla)  \vec{v} = - \frac{1}{\Sigma} \nabla P - \nabla \Phi,
\eeq
with $\Phi$ denoting the gravitational potential of the central star and $P$ denoting the pressure.
The interaction between gas and dust due to aerodynamical friction is implemented as a drag force
\beq\label{eq:gasdrag}
\vec{f}_{\mathrm{aero}} = - \Sigma_{d} \Omega_{K}(r) \mathrm{St}^{-1} (\vec{v}_{d} -\vec{v}_{g}),
\eeq
with $\Omega_K^2 = GM_\star /r^3$ describing the Keplerian orbital frequency and $\mathrm{St}$ being the Stokes number, see \cite{Weidenschilling1977}. 
The Stokes number is the ratio of the stopping time $t_{\mathrm{stop}}$ and the turn over time of the largest turbulent eddy $t_{\mathrm{turnover}}$. 
The later may be written as $t_{\mathrm{turnover}} = \Omega_{\mathrm{K}}^{-1}$ (see \citealt{Cuzzi2001}) and assuming the Epstein drag law, an isothermal volumentric gas density profile with the gas density in the midplane $\rho_g = \Sigma_g \Omega_k / \sqrt{2 \pi} c_s$ and spherical particles, we can write the Stokes number as
\beq
\mathrm{St} = \frac{t_{\mathrm{stop}}}{t_{\mathrm{turnover}}} = \frac{\pi}{2} \frac{a \rho_s }{\Sigma_g},
\eeq
with $a$ denoting the particle radius and $\rho_s$ the particle density. In this work we always use 1 $g/cm^3$ for the particle density $\rho_s$.

Combining Eq. \ref{eq:gasevo} and \ref{eq:gasdrag} in polar coordinates leads to the radial gas velocity evolution:
\beq
\begin{aligned}
& \frac{\partial}{\partial t}v_{g,r} +  v_{g,r} \frac{\partial}{\partial r} v_{g,r} -  \frac{v_{g,\phi}^{2} }{g,r}  + \frac{v_{g,\phi} }{r}  \frac{\partial}{\partial \phi} v_{g,r}  \\
=& -\frac{1}{\Sigma_{g}}  \frac{\partial}{\partial r} P  -  r \Omega_{K}^{2} - \frac{\vec{f}_\mathrm{aero}\cdot \vec{e}_{r}}{\Sigma_{g}},
\end{aligned}
\eeq
and for the azimuthal gas velocity component we obtain:
\beq
\begin{aligned}
& \frac{\partial}{\partial t} v_{g,\phi} + \frac{1}{r} v_{g,\phi}  \frac{\partial}{\partial \phi} v_{g,\phi} + \frac{ v_{g,r}}{r} v_{g,\phi} + v_{g,r}  \frac{\partial}{\partial r} v_{g,\phi}  \\
=& - \frac{1}{\Sigma_{g}}   \frac{1}{r} \frac{\partial}{\partial \phi} P - \frac{\vec{f}_\mathrm{aero}\cdot \vec{e}_{\phi}}{\Sigma_{g}},
\end{aligned}
\eeq
with $\vec{e}_r$ and $\vec{e}_\phi$ being the radial and azimuthal unit vectors.
The corresponding equations for the dust can be written as
\beq
\begin{aligned}
& \frac{\partial}{\partial t}v_{d,r} +  v_{d,r} \frac{\partial}{\partial r} v_{d,r} -  \frac{v_{d,\phi}^{2} }{r}  + \frac{v_{d,\phi} }{r}  \frac{\partial}{\partial \phi} v_{d,r} \\
=&  -  r \Omega_{K}^{2} + \frac{\vec{f}_\mathrm{aero}\cdot \vec{e}_{r}}{\Sigma_{d}},
\end{aligned}
\eeq
and
\beq
\begin{aligned}
& \frac{\partial}{\partial t} v_{d,\phi} + \frac{1}{r} v_{d,\phi}  \frac{\partial}{\partial \phi} v_{d,\phi} + \frac{ v_{d,r}}{r} v_{d,\phi} + v_{d,r}  \frac{\partial}{\partial r} v_{\phi}  \\
= & \frac{\vec{f}_\mathrm{aero}\cdot \vec{e}_{\phi}}{\Sigma_{d}}.
\end{aligned}
\eeq

The evolution of pressure is obtained by solving the adiabatic energy equation. If the energy conservation is solved, the corresponding pressure can be calculated after each time step from the total energy of the adiabatic gas:
\beq
E = \frac{P}{1-\gamma} + \frac{1}{2} \Sigma_g \vec{v}_g^{2},
\eeq
with $\gamma = 1.4$. Hence the pressure is updated according to the energy conservation
\beq
P = (1-\gamma)\cdot \left (E - \frac{1}{2}  \Sigma_g  \vec{v}_g^{2} \right).
\eeq

We use the fluid approximation for the gas and a pressure-less fluid model for the dust content. 
This assumption is justified because in our model the dust particles stay small, within the Epstein drag regime. 
Consequently particles within one grid cell have nearly the same properties and can therefore be computationally modeled as a fluid.
However particles in the Epstein regime mainly interact with the gas molecules and not with other dust particles, hence the pressure-less fluid assumption has to be used.

The conservative form of this system of coupled equations is solved by the RoSSBi code using a well-balanced finite volume method, see \cite{Surville2015} and \cite{Surville2016}. The time integration is based on a second order Runge-Kutta scheme and the flux reconstruction is third order in space using parabolic interpolation. The numerical flux of the gas fluid are obtained by an exact Riemann solver, and the ones of the dust fluid are obtained by a pressure-less Roe solver used in \cite{Paardekooper2006a}. 

The boundary conditions implemented in the code RoSSBi are based on zero gradient conditions, where ghost cell variables are reconstructed to follow the steady state profiles of the disk. These free conditions account for radial flux of gas and dust. However, in order to keep stability of the dust fluid (in particular at the outer disk boundary), only the dust density is damped toward the initial profile.

Additionally, the gravity of an embedded planet is implemented in the code RoSSBi. In some of the runs presented in this study, a Jupiter mass planet is orbiting on a circular orbit $\vec{r}_p$ around the star. As a simplification, the center of mass of the star/planet system is kept at the star center, which is the origin of the reference frame.

The field of gravitational force exerted by the planet of mass $M_p$ on the disk is given by
\beq
	\vec{g}_{pla}(\vec{r}) = -\frac{G M_p}{{\left[|\vec{r} - \vec{r}_p| + l(r_p) \right]}^{3}} \left(\vec{r} - \vec{r}_p \right),
\eeq
where the gravity is modified using a well known softening length $l(r_p) = 0.6 H_0(r_p)$ for a planet potential. The isothermal disk scale height at the planet orbit is $H_0(r_p)=\left[P_0(r_p)/\Sigma_0(r_p)\right]^{1/2}/\Omega_K(r_p) = c_{s}(r_0)/\Omega_r(p_0)$, with $c_s$ denoting the sound speed. The actual profiles of the gas background pressure and density will be given in Section \ref{s:IC}. 
Finally, the mass of the planet is loaded directly from the beginning of the run, which has no critical influence on the evolution of the disk later on.
The drag interaction between dust and gas fluids is solved using an implicit method described in \cite{Surville2018}. 
Finally, simulations including a planet are done using a thermal relaxation term in the energy equation to avoid shock heating of the disk.
This additional source term is computed implicitly.
Thus, the time step is determined by the CFL condition, with a factor 0.5 needed by the parabolic reconstruction of the RoSSBi scheme.
For this simulations, except in the ZHU and DZHU simulations (see Section \ref{s:ATHENA}), we use an average grid size of 0.0244 AU in radius and an average aspect ratio of 0.285. 
The usage of a cell aspect ratio (hereafter CAR) of $\sim$0.25-0.3 has been tested and robustly confirmed in several papers using the RoSSBi code (\citealt{Surville2015}, \citealt{Surville2016, Surville2018}) for the evolution of vortices and also the convergence at higher resolutions for these CAR as been tested during the preparation of the paper.
Therefore we can resolve the Hill radius, in the simulations containing a planet, with 78 grid cells in radius and 69 cells in azimuth.

\subsection{Dust evolution}\label{s:Dustevo}
We base our approach for modeling dust sizes on the so-called two population algorithm proposed by \cite{Birnstiel2012}. 
The idea behind that model was to reproduce the general pattern of dust surface density evolution obtained with the dust coagulation code of \cite{Birnstiel2010} at much reduced computational cost, without actually solving the dust coagulation equation. 
In each grid cell the dust size $a_{\rm{max}}$ is chosen to represent the full dust size distribution, basing on a semi-analytical approach.
Making the right choice of $a_{\rm{max}}$ is possible thanks to the comprehensive understanding of the processes governing dust evolution presented by \cite{Birnstiel2012}.

The original dust evolution model of \cite{Birnstiel2012} was one dimensional.
Here we extend this method to two dimensions, by applying the sub-grid dust growth and fragmentation prescription and keeping the original advection scheme of the RoSSBi code. 
In this model, the representative dust particle size, determined with the help of local gas and dust variables (see Eq. \ref{eq:afrag} and Eq. \ref{eq:initgrowth}), is calculated before each Runge-Kutta time step and in each single grid cell.
This representative grain size is then used to compute the aerodynamical friction source terms in the RoSSBi code.
As the required calculation is local to each grid cell, this subgrid method does not change the parallelization model of the hydrodynamical method.\\
The representative size $a_{\rm{max}}$ is found in each cell by comparing the maximum aggregate size that could be obtained taking into account various physical processes: dust growth (Eq.~\ref{eq:initgrowth}), fragmentation (Eq.~\ref{eq:afrag} and Eq. \ref{eq:adf}) and the loss of large aggregates due to radial drift (Eq.~\ref{eq:firstlimit}). 
We pick the smallest of these sizes as the representative size characterizing the local dust population.
In what follows we describe how we model such individual processes.

The growth timescale can be written as
\beq\label{eq:growthtime}
\tau_{\text{grow}} = \frac{a}{\dot{a}} \approx \frac{1}{Z \Omega_K},
\eeq
with $Z$ standing for the vertically integrated dust to gas ratio. In the inner parts of protoplanetary disk, this timescale is typically faster than the global dust redistribution timescale (\citealt{Birnstiel2012}, \citealt{Drcazkowska2016}). In such a case, the dust size distribution is governed by a coagulation-fragmentation equilibrium. 
Since the impact speeds of dust aggregates increase with their size, there is a maximum size that can be obtained by dust growth before it is halted by fragmentation. If the dominant source of the impact speeds is turbulence, the representative size is
\beq\label{eq:afrag} 
a_{\text{frag}} = \frac{f_{\rm{f}}}{3}  \frac{2 \Sigma_g}{\pi \rho_{s}} \frac{u_{f}^{2}}{\alpha c_{s}^{2}} ,
\eeq 
where $f_{\rm{f}}$ is an order of unity constant, $\alpha$ is the turbulence parameter (see \citealt{Shakura1973}), $\rho_s$ is the internal density of dust particles and $u_f$ denotes the threshold fragmentation speed, which we set to $u_f=10$~m~s$^{-1}$.
      The turbulence that drives impact speeds has typical eddy overturn timescales comparable to a typical stopping time of dust grains, these correspond, in our models, to eddies of roughly the size of 1 grid cell.
 This implies that we could not resolve them and therefore the $\alpha$ parameter is describing a sub-grid turbulence. 
 In this paper we chose $\alpha = 10^{-3}$ because it is the standard value used in dust coagulation models.
The maximum particle size possible to obtain with respect to the impact speeds triggered by the differential drift can be written as
\beq\label{eq:adf}
a_{\text{df}} = \frac{2\Sigma_{g}}{\pi \rho_s} \frac{ u_f V_k}{c_s^2(1-N)} \left|\frac{d\ln{P}}{d\ln{r}}\right|^{-1},
\eeq
where N is defined as a typical ratio between the Stokes numbers of two colliding particles and $V_k$ the Keplerian velocity.
Following \cite{Birnstiel2012}, we use $N = 0.5$, since this gives the best fit to complete models.

In the outer parts of the protoplanetary disk, particles growth timescale becomes longer than the radial drift timescale. 
Since in our approach the advection of dust does not have a direct effect on dust size, this process cannot be explicitly modeled. 
Therefore we need to include a drift limit, which takes this effect explicitly in to account.
Leaving such a limit out of consideration, would lead to an over-prediction of the grain size because if the drift timescale is shorter than the growth timescale, dust grains should be removed by the radial drift before they grow to the size limited by fragmentation.
The maximum size that can be kept at a given orbital distance before it would be removed by the drift can be written as
\beq\label{eq:firstlimit}
a_{\rm{drift}} = {\rm{f_d}} \frac{2\Sigma_{g}}{\pi \rho_s}\frac{V_{k}^2}{c_{s}^2}\left|\frac{d\ln{P}}{d\ln{r}}\right|^{-1},
\eeq
with $f_d$ being the numerical constant for drift.
Moreover, we use an initial growth limit
\beq \label{eq:initgrowth}
a_{\text{ini}} =  a_{0} \cdot \exp \left( \frac{t}{\tau_{grow}} \right),
\eeq
with $a_0$ denoting the initial particle size, and the growth timescale $\tau_{\text{grow}}$ (described by Eq.~\ref{eq:growthtime}). 
This limit takes into account that the growth timescale significantly increases with the orbital distance, so the large particles occur in the outer part of the disk much later than in its inner part (\citealt{Birnstiel2012} called this effect "the delayed drift effect" and \citealt{Lambrechts2014} "the pebble formation edge").

The algorithm of \cite{Birnstiel2012} considered two characteristic sizes of dust particles in each cell: the minimum and the maximum size. In our code, we are restricted to a single size per grid cell. Hence, the smallest dust size in each cell is neglected and the whole dust surface density is assumed to be generated by the maximum sized particles. 
This is a good estimate if the particle size is limited by radial drift, because $97\%$ of the dust surface density is determined by the mass of the $a_{\text{max}}$ sized particles.
As for the turbulence barrier, 75$\%$ of the dust mass is in the maximum sized particles.

\subsection{Initial conditions}\label{s:IC}
In this work we chose the Minimum Mass Solar Nebula (hereafter MMSN) model, see \cite{Hayashi1981}, as the initial dust surface density model, but the configuration of the RoSSBi code allows to change it to another model in the future. 
This particular choice is justified with the MMSN being a well studied model for which it is relatively simple to find simulations to be compared with this work. 
Within the MMSN model the initial gas surface density can be written as:
\beq
\Sigma_g(r>2.7\text{AU})= 1700 \left( \frac{r}{\text{AU}} \right)^{-3/2} \left[\frac{\text{g} }{\text{cm}^{2}}\right].
\eeq
The assumption of an adiabatic ideal gas combined with a power law surface density profile leads to the following simple 
form of the pressure profile of the disk (see also \citealt{Surville2016}):
\beq
P = P_{0} \cdot  \left( \frac{r}{r_{0}} \right)^{-2},
\eeq
with $r_0$ the reference radius, which corresponds to the planet location in the runs with a planet ($r_0=10$ AU).
In order to normalize the pressure and set the temperature profile we normalize the disk scale height to $H_0(r_0)=0.05 \: r_0$. 
Supplementary information can be found in \cite{Surville2016}.
The domain size ranges always from 5 to 30 AU and simulations are stopped after 400 orbits at 10 AU. 
For the simulations with an embedded planet we consider always a Jupiter mass planet at 10 AU.
We note that in runs containing a planet we apply a fast cooling function to the energy equation which makes the simulations effectively isothermal. 
The fast cooling function can be written as  
\beq
f_{cool} =  -\Sigma_g (T-T_0) \frac{\Omega_k}{\tau_{cool}},
\eeq
with $\tau_{cool}$ denoting the cooling constant, $T$ the temperature and $T_0$ the reference temperature at the planet position. In this study we choose $\tau_{cool} = 10^{-4}$.

\section{Results}\label{s:masterresults}
The major objective of this work is to compare the state of the art models, which are based on a fixed dust size approximation, to the new sub-grid model, employing a variable dust size. 
Additionally, if qualitative differences between the two methods are found, we want to provide an interpretation of the results.
Thus, we restricted the work presented in this paper to relatively simple setups, which clearly show the differences between the two methods.
The impact of our results on the interpretation of protoplanetary disk observations will be the subject of future papers.
A complete list of all simulations carried out with the RoSSBi code employing a fixed dust size 
can be viewed in Table \ref{tab:rossbi} along with the list of the corresponding runs allowing for a variable 
dust size according to our sub-grid model is included in Table \ref{tab:dustevo}. 

\begin{deluxetable*}{cccccccc}
\tablecaption{Fixed dust sized simulations\label{tab:rossbi}}
\tablecolumns{8}
\tablehead{
\colhead{Identifier}&
\colhead{Planet} &
\colhead{$a_0$} & 
\colhead{Dust}&
\colhead{$\mathrm{St}(r_0)$} &
\colhead{$\Sigma_d / \Sigma_g$} & 
\colhead{Resolution} &
\colhead{Heat} \\
\colhead{} & \colhead{} &
\colhead{[cm]} & 
\colhead{evolution}&
\colhead{} & \colhead{} & 
\colhead{[R,$\Theta$] } & 
\colhead{transfer}
}
\startdata
RN32&NO&3&NO&$8.4 \cdot 10^{-2}$&$10^{-2}$&1024x1024&Adiabatic\\
RN1-42&NO&$10^{-4}$&NO&$2.8 \cdot 10^{-6}$&$10^{-2}$&1024x1024&Adiabatic\\
RY32&YES&3&NO&$8.4 \cdot 10^{-2}$&$10^{-2}$&1024x1024&Isothermal\\
RY1-12&YES&$10^{-1}$&NO&$2.8 \cdot 10^{-3}$&$10^{-2}$&1024x1024&Isothermal\\
RY1-22&YES& $10^{-2}$&NO&$2.8 \cdot 10^{-4}$&$10^{-2}$&1024x1024&Isothermal\\
RY1-42&YES&$10^{-4}$&NO&$2.8 \cdot 10^{-6}$&$10^{-2}$&1024x1024&Isothermal\\
ATH&YES&$2$&NO&$1.76 \cdot 10^{-2}$&$10^{-2}$&282x1024&Isothermal\\
\enddata
\tablecomments{Overview of all simulations done with the RoSSBi code, assuming a fixed dust size. The identifiers of the individual simulations are derived from the simulation setup [e.g. 'RY' for 'RoSSBi' with planet, the particle size, e.g. $10^{-4}$cm is written as '1-4', and the gas to dust ratio, e.g. $10^{-2}$ is simply written as 2. A combination of this identifiers will lead to 'RY1-42']. The table also displays the Stokes number at the reference radius ($r_0 = 10$ AU), the resolution and also which heat transfer is applied in each simulation.}
\end{deluxetable*}

\begin{deluxetable*}{cccccccc}
\tablecaption{Variable dust size simulations\label{tab:dustevo}}
\tablecolumns{8}
\tablehead{
\colhead{Identifier}&
\colhead{Planet} &
\colhead{$a_0$} & 
\colhead{Dust}&
\colhead{$\mathrm{St}(r_0)$} &
\colhead{$\Sigma_d / \Sigma_g$} & 
\colhead{Resolution} &
\colhead{Heat} \\
\colhead{} & \colhead{} &
\colhead{[cm]} & 
\colhead{evolution}&
\colhead{} & \colhead{} & 
\colhead{[R,$\Theta$] } & 
\colhead{transfer}
}
\startdata
DN32&NO&3&YES&$8.4 \cdot 10^{-2}$&$10^{-2}$&1024x1024&Adiabatic\\
DN1-42&NO&$10^{-4}$&YES&$2.8 \cdot 10^{-6}$&$10^{-2}$&1024x1024&Adiabatic\\
DY32&YES&3&YES&$8.4 \cdot 10^{-2}$&$10^{-2}$&1024x1024&Isothermal\\
DY1-12&YES&$10^{-1}$&YES&$2.8 \cdot 10^{-3}$&$10^{-2}$&1024x1024&Isothermal\\
DY1-22&YES& $10^{-2}$&YES&$2.8 \cdot 10^{-4}$&$10^{-2}$&1024x1024&Isothermal\\
DY1-42&YES&$10^{-4}$&YES&$2.8 \cdot 10^{-6}$&$10^{-2}$&1024x1024&Isothermal\\
DATH&YES&$2$&YES&$1.76 \cdot 10^{-2}$&$10^{-2}$&282x1024&Isothermal\\
\enddata
\tablecomments{A summary of all simulations including effects of dust coagulation and fragmentation. The identifiers of the individual simulations follow the same scheme as described in Table \ref{tab:rossbi}. The table specifies the Stokes number at the reference radius ($r_0 = 10$ AU) at the beginning of each simulation, the resolution and the heat transfer type used.}
\end{deluxetable*}

\subsection{Results of simulations without planet}\label{s:noplanet}

\begin{figure*}
\plotone{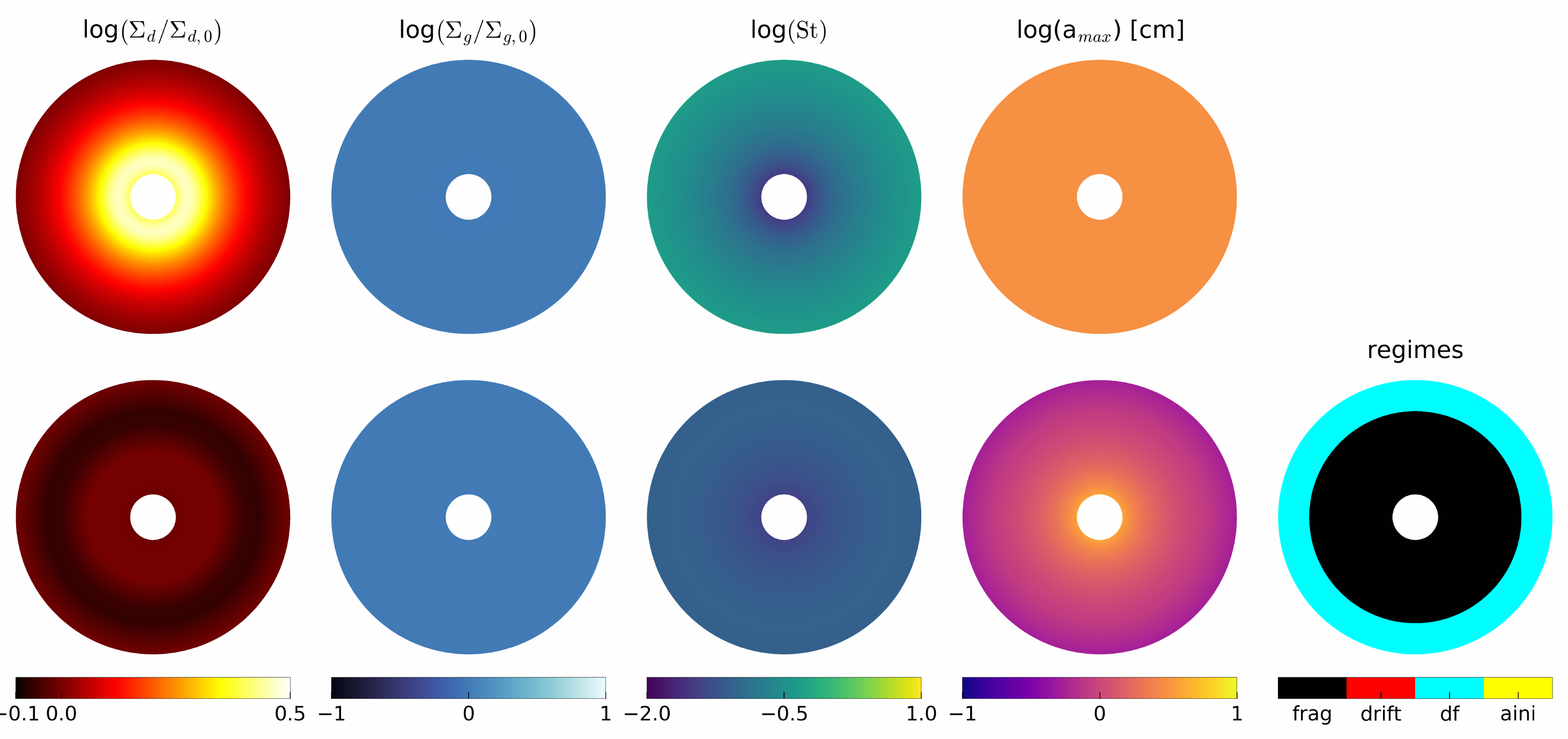}
\caption{{\it The leftmost two upper panels:} Dust and gas surface densities normalized by their initial values after 400 orbits obtained in simulations with dust size fixed at 3 cm (RN32) and an initial MMSN gas disk model. 
The other two upper panels show the Stokes number and the particle size in cm.
{\it Lower panels:} Results of the corresponding run with realistic dust size (DN32). The panels show the normalized dust and gas surface densities, Stokes number, particle size, and the process that determines the particle size, respectively.}
\label{fig:DN32}
\end{figure*}
\begin{figure*}  
\plotone{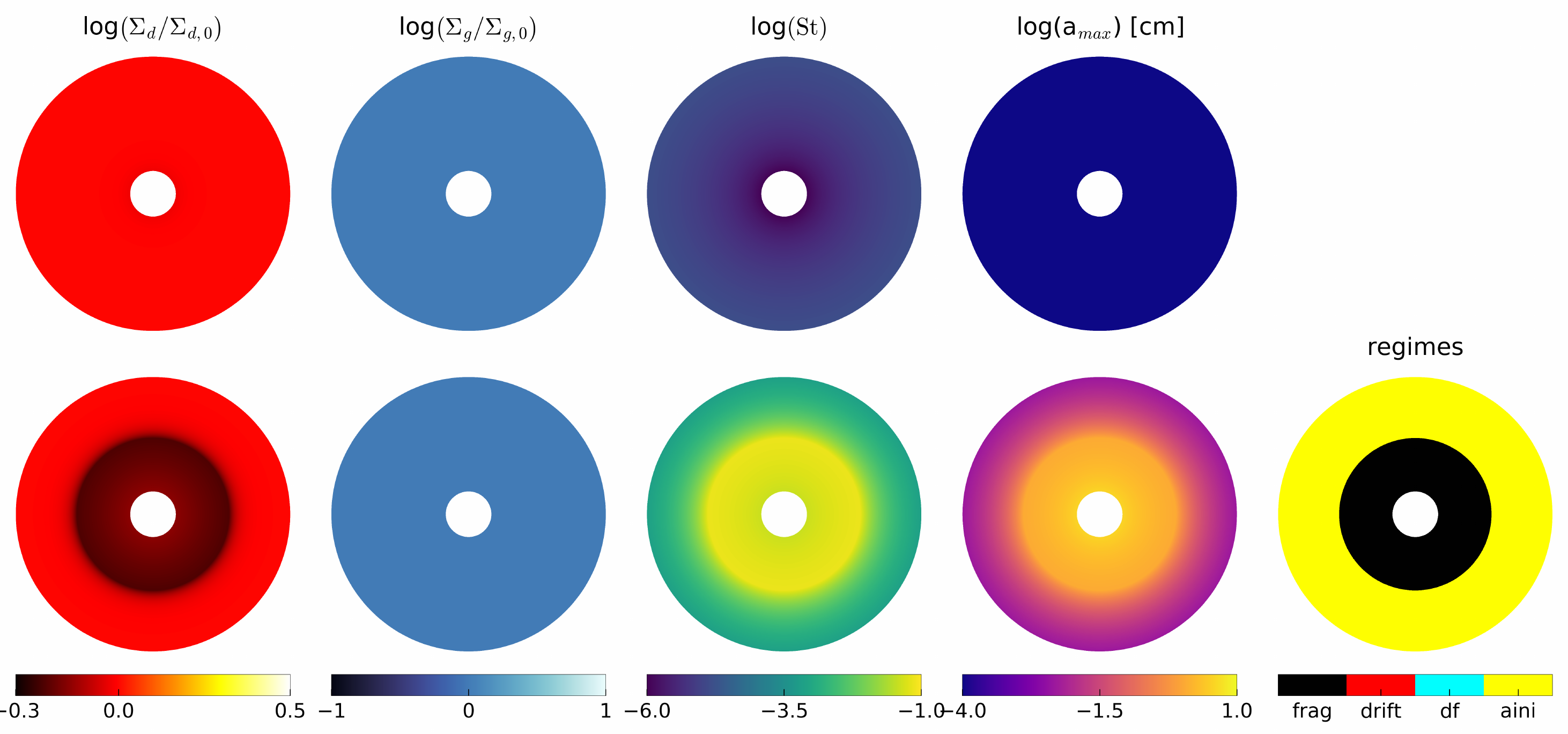}
\caption{{\it The leftmost two upper panels:} Dust and gas surface densities normalized by their initial values after 400 orbits obtained in simulations with dust size fixed at $10^{-4}$~cm (RN1-42) and an initial MMSN gas disk model. The other two upper panels show the Stokes number and the particle size in cm.
{\it Lower panels:} Results of the corresponding run with dust evolution enabled (DN1-42). The panels show the normalized dust and gas surface densities, Stokes number, particle size, and the process that determines the particle size, respectively.}
\label{fig:DN1-42}
\end{figure*}

Many hydrodynamic grid codes rely on assumption that dust size (or alternatively, the Stokes number) is fixed across the simulated domain. 
Often this size is is set to represent large grains, which are expected to grow in some regions of protoplanetary disks. 
The upper panel in Fig. \ref{fig:DN32} presents the results of such a simulation, with the dust size fixed to 3 cm (RN32). 
The gas phase exhibits no significant evolution past the initial condition of this run. 
But the dust component is axisymmetrically pilling up in the inner part of the disk. 
This is a result of the inward drift and the inflow at the outer boundary. 
Since the particle size is fixed the Stokes number of particles increases with radial distance. 
Additionally the radial drift velocity increases with the Stokes number (until $\mathrm{St} <1$), hence particles in the outer part of the domain drift faster than those in the inner part, causing the increase of dust density at the inner edge of the domain.

The lower panel of Fig. \ref{fig:DN32} presents results of the corresponding run
including our sub-grid model for dust evolution (DN32). 
For consistency, we initialized this simulation with 3 cm grains, 
which means that the initial growth stage is by-passed as dust size is 
immediately adjusted to the maximum possible so that at each radial distance 
fragmentation and radial drift become the governing mechanisms.
The figures clearly illustrates that the
pattern of surface density evolution is quite different from that in the fixed size simulation. 
Particularly, no pile-up of dust in the inner disk arises. 
The dark annulus marks depletion of dust that develops at the boundary between the regions dominated by fragmentation, which is triggered by turbulence and radial drift.

The first conclusion to be drawn is that here is a qualitative change in the dust evolution 
pattern depending on whether we assume fixed or variable dust size even if we start from large
grains in both cases. We can now proceed to compare the dust evolution pattern when
we start from small grains and consider the initial growth stage. 
Figure~\ref{fig:DN1-42} presents results of two runs (RN1-42 and DN1-42) that begin with 
dust grains having a size of 1~$\mu$m, which is typically assumed 
as an initial size in works modeling dust growth in protoplanetary disks (\citealt{Windmark2012}). 
In the upper panel of Fig. \ref{fig:DN1-42}, this size is fixed throughout the evolution and thus the dust surface density practically does not evolve as the small grains stay well-coupled to the gas. 
The bottom panel of Fig. \ref{fig:DN1-42} displays the inside-out evolution pattern commonly obtained in one dimensional dust growth models (\citealt{Birnstiel2012}, \citealt{Krijt2016}). 
The particles grow faster at smaller orbital distances, where the Keplerian frequency is higher (see Eq.~\ref{eq:initgrowth}). 
Those particles then decouple form the gas disk and drift inward, before the particles in the outer part of the domain manage to reach their maximum size.

\subsection{Results of simulations including planet}\label{s:masterplanet}

\begin{figure*}
\plotone{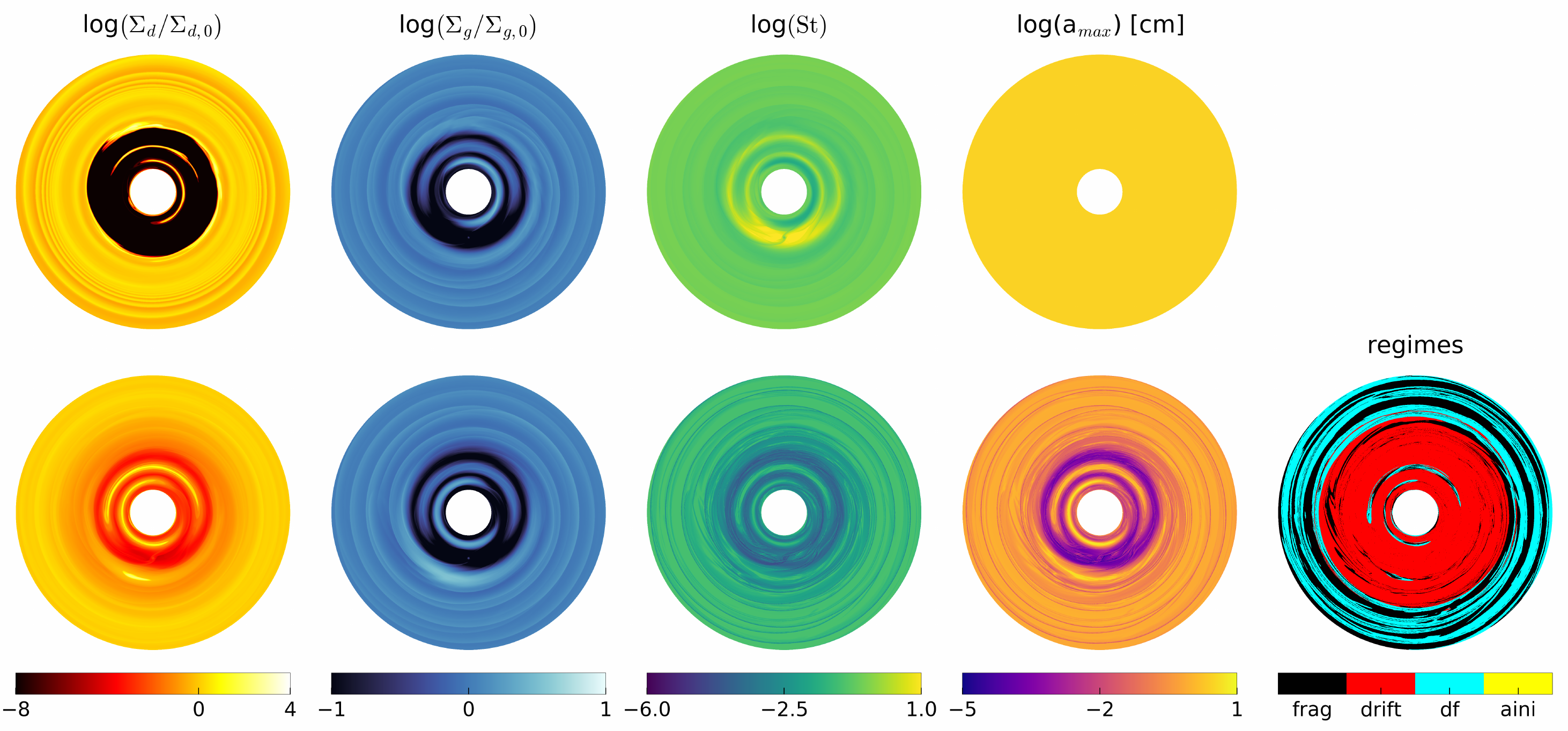}
\caption{{\it The two righmost upper panels:} Dust and gas densities normalized by their initial values after 400 orbits obtained in simulations with dust size fixed at 3 cm and a Jupiter mass planet placed at 10~AU (RY32). 
The other two upper panels show the Stokes number and the constant particle size in cm.
{\it Lower panels:} Results of the analogical run with realistic dust size (DY32). The panels show the normalized dust and gas surface densities, Stokes number, particle size, and the process that determines the particle size, respectively.}
\label{fig:DY32}
\end{figure*}

Figure \ref{fig:DY32} illustrates the results of two runs including Jupiter mass planet at 10~AU: one with 
the dust size fixed to 3 cm (RY32, \textit{top panel}) and one including our dust evolution model starting from grains having a size of 3 cm (DY32, \textit{bottom panel}). 
Adding a planet to the simulations effects not only the dust component but also the gas.
This is explained by the additional gravitational potential of the planet as well as by the back reaction of the dust fluid to the gas component. 
Our results for the gas surface densities and gap opening are comparable with previous studies, e.g \cite{Rosotti2016}.
As seen in the upper panel of Fig. \ref{fig:DY32}, the planet induced pressure bump stops the migration of particles and therefore a ring structure behind the planet is formed (see e.g. \citealt{Paardekooper2004, Paardekooper2006a} , \citealt{ValBorro2007}, \citealt{Fouchet2010}, \citealt{Gonzalez2012}, and \citealt{Zhu2012, Zhu2014}). 
Particles in the inner part of the planetary orbit can simply drift towards the central star and vanish from the simulated disk.
Since in the fixed sized RY32 simulation the 3 cm particles have high drift velocities, particles from the outer part of the disk arrive at the pressure bump quickly. 

The bottom panel of Fig. \ref{fig:DY32} shows results of the run with a planet and the variable dust size algorithm (DY32). 
The evolution of the disk is less dramatic, hence slower than the evolution of the fixed size simulation.
This is because our dust size routine returns sizes that are, with the exception of very inner part of the domain, significantly lower than the 3 cm used in the fixed size case. 
The presence of the planet triggers strong pressure gradients and thus particles in the disk are now also controlled by the radial drift.
In the variable size simulation we do not obtain the pronounced dust annulus at the outer edge of planetary gap as observed in the fixed size simulation. 
Also, the depletion of inner disk is significantly weaker.

\subsection{Detailed comparison between fixed and variable sized dust grains}\label{s:mastercomparison}
The first part of this section exemplified the results of each individual method. 
In order to compare them in a more qualitative approach, in this subsection we investigate histograms of azimuthally averaged and normalized values for the dust and gas surface density component.

In the absence of the planet, the azimuthal averaging is straightforward and does not reduce the amount of information. 
However, using one analysis tool for all simulations and applying it to axisymmetric solutions may be used as a test for its accuracy. 
Because the simulations start with the same conditions, a change in the results is a product of the different particle sizes.

\begin{figure*}
\plotone{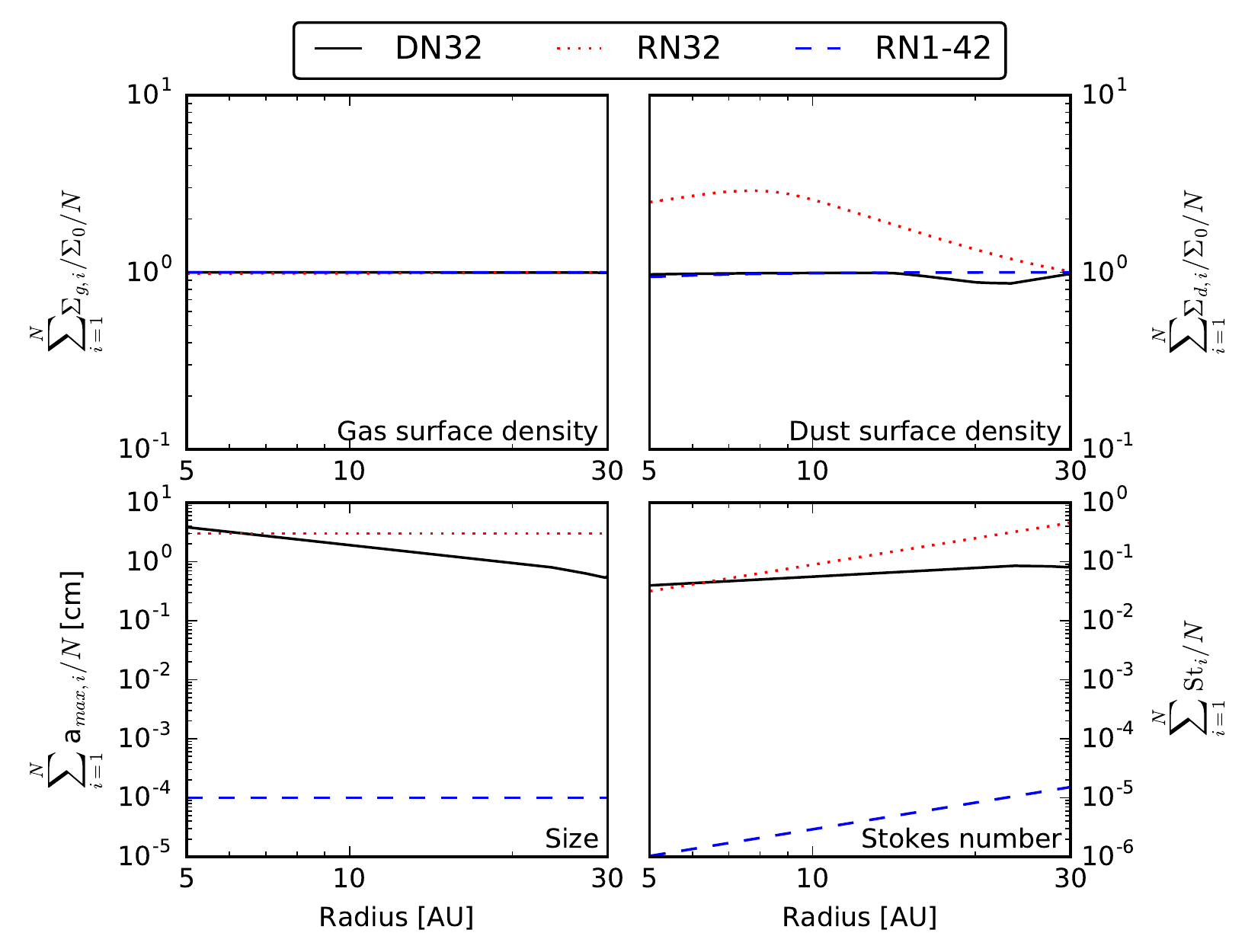}
\caption{A comparison of the fixed dust size simulation with 3 cm RN32 (red dotted line), $10^{-4}$ cm RN1-42 (blue dashed line) and a simulation with dust evolution starting with 3 cm sized particles DN32 (solid black line) after 400 orbits. In all three simulations, the gas component is nearly unchanged in the azimuthally averaged and normalized gas surface density (\textit{top left panel}). A different result is observed for the dust component (\textit{top right panel}). The two bottom panels show the averaged particle size in cm (\textit{left panel}) as well as the averaged Stokes number (\textit{right panel}).}
\label{fig:compA_noplanet}
\end{figure*}

The simulations with a 3 cm fixed dust size (RN32), a $10^{-4}$ cm fixed dust size (RN1-42) and a variable dust size, starting with 3 cm sized particles (DN32), without a planet show no significant change of the gas surface density profiles, as presented in Fig. \ref{fig:compA_noplanet}.
This is expected since a change of the gas surface density occurs only via the backreaction from dust to gas.
Because of the low dust-to-gas ratio, the impact of the backreaction during the simulated timescales in an axisymmetric simulation should be negligible and therefore the changes in the gas surface density are insignificant.

On the other hand, the dust surface density evolution is directly affected by the particle size due to radial drift. 
As shown, in Fig. \ref{fig:compA_noplanet}, the large dust sizes lead to a pile up in the inner part of the disk. 
In contrast, the dust surface density for the small particles is nearly unchanged, because their drift velocity is much lower and the evolution time is too short to allow for a significant drift.

\begin{figure*}
\plotone{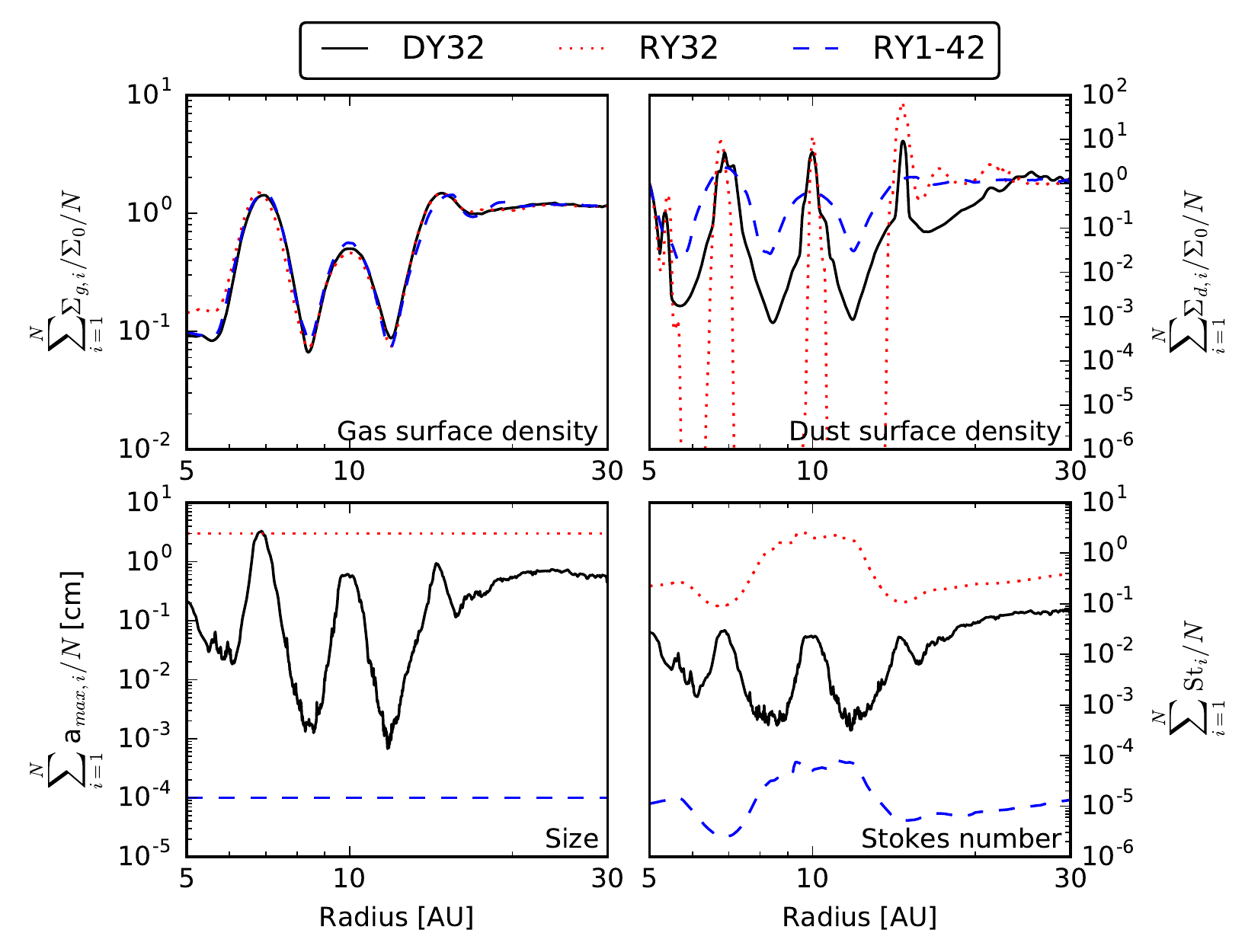}
\caption{Comparison of a fixed dust size simulation with 3 cm sized dust particles RY32 (red dotted line), a variable dust size simulations starting form 3 cm sized particles DY32 (solid black line) and a fixed sized simulation containing $10^{-4}$ cm sized particles RY1-42 (dashed blue line) after 400 orbits. All simulations contain a Jupiter sized planet at 10AU and a MMSN gas disk model. The bottom two panels show the averaged particle size in cm (\textit{left panel}) as well as the averaged Stokes number (\textit{right panel}).}
\label{fig:compA_planet}
\end{figure*}

The azimuthally averaged outcomes of the RY32, RY1-42 and DY32 runs including the planet, after 400 orbits, can be seen in Fig. \ref{fig:compA_planet}. 
Only a slight difference among the three gas surface densities is visible.
Thus the impact of differently sized particles appears to be negligible for the gas evolution, since the overall shape and position of the curve is the same for all three simulations.
The dust surface density, however exhibits a completely different picture. 
In case of the large fixed sized particles the disk gets almost depleted, since mainly dust particles, which are trapped in the pressure bump behind the planet, remain in the disk. 
Particles from the outer edge of the disk migrate to this barrier and the surface density peaks at this position within the disk. 
On the other hand, the small particles are well coupled to the gas and, as a result, they migrate slowly and are not trapped by the pressure bump. 
Comparing the two fixed sized simulations RY32 and RY1-42 (Fig. \ref{fig:compA_planet}) yields an enormous difference in the final dust surface density.

\begin{figure*}
\plotone{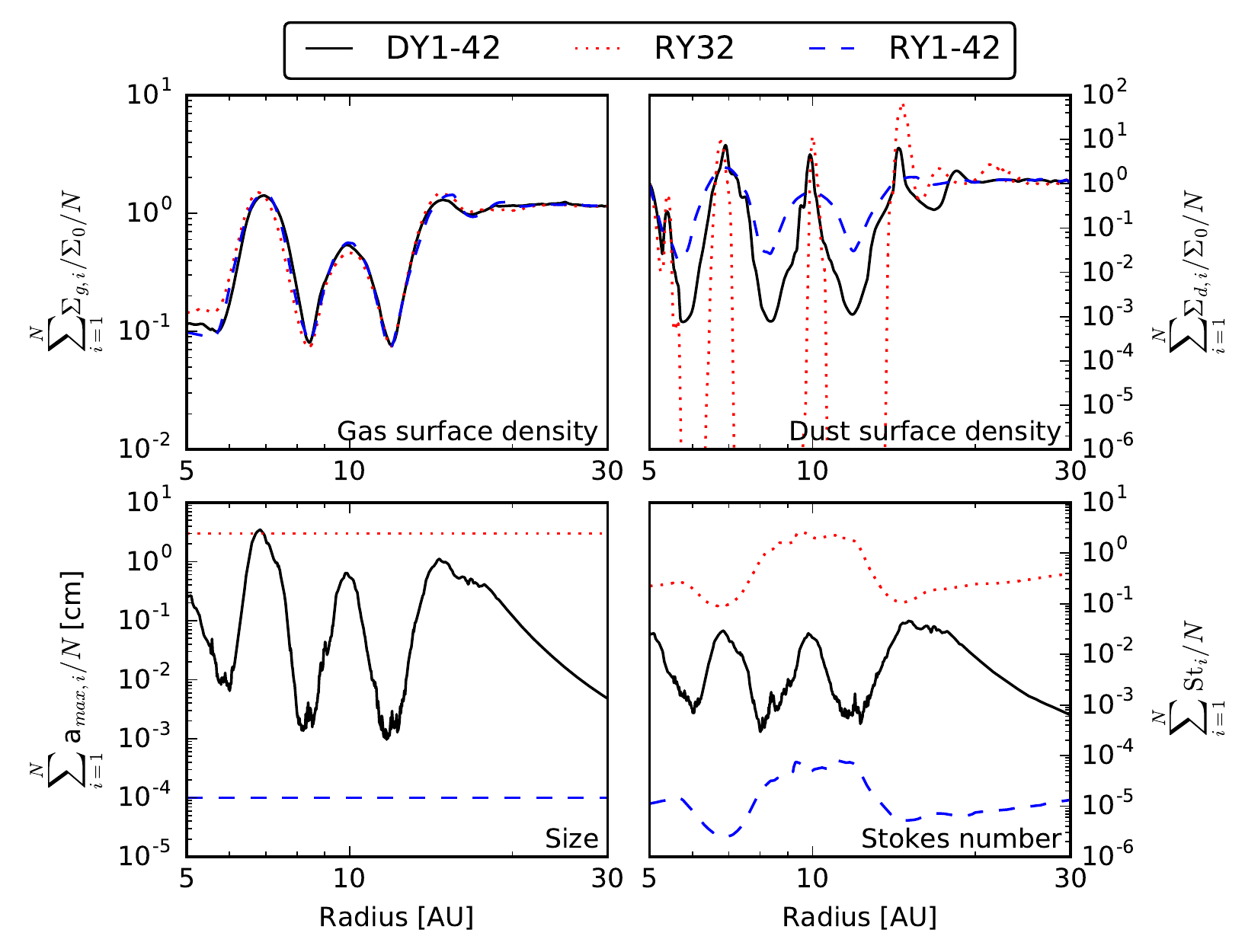}
\caption{ Comparison of a fixed dust size simulation with 3 cm sized dust particles RY32 (red dotted line), a variable dust size simulations starting form $10^{-4}$ cm sized particles DY1-42 (solid black line) and a fixed sized simulation containing $10^{-4}$ cm sized particles RY1-42 (dashed blue line) after 400 orbits. All the simulations contain a Jupiter sized planet at 10AU and a MMSN gas disk model. The azimuthally averaged and normalized gas and dust surface density can be seen in the upper two panels. The bottom two panels show the averaged particle size in cm (\textit{bottom left panel}) as well as the averaged Stokes number (\textit{bottom right panel}).}
\label{fig:compB_planet}
\end{figure*}

\begin{figure*}
\plotone{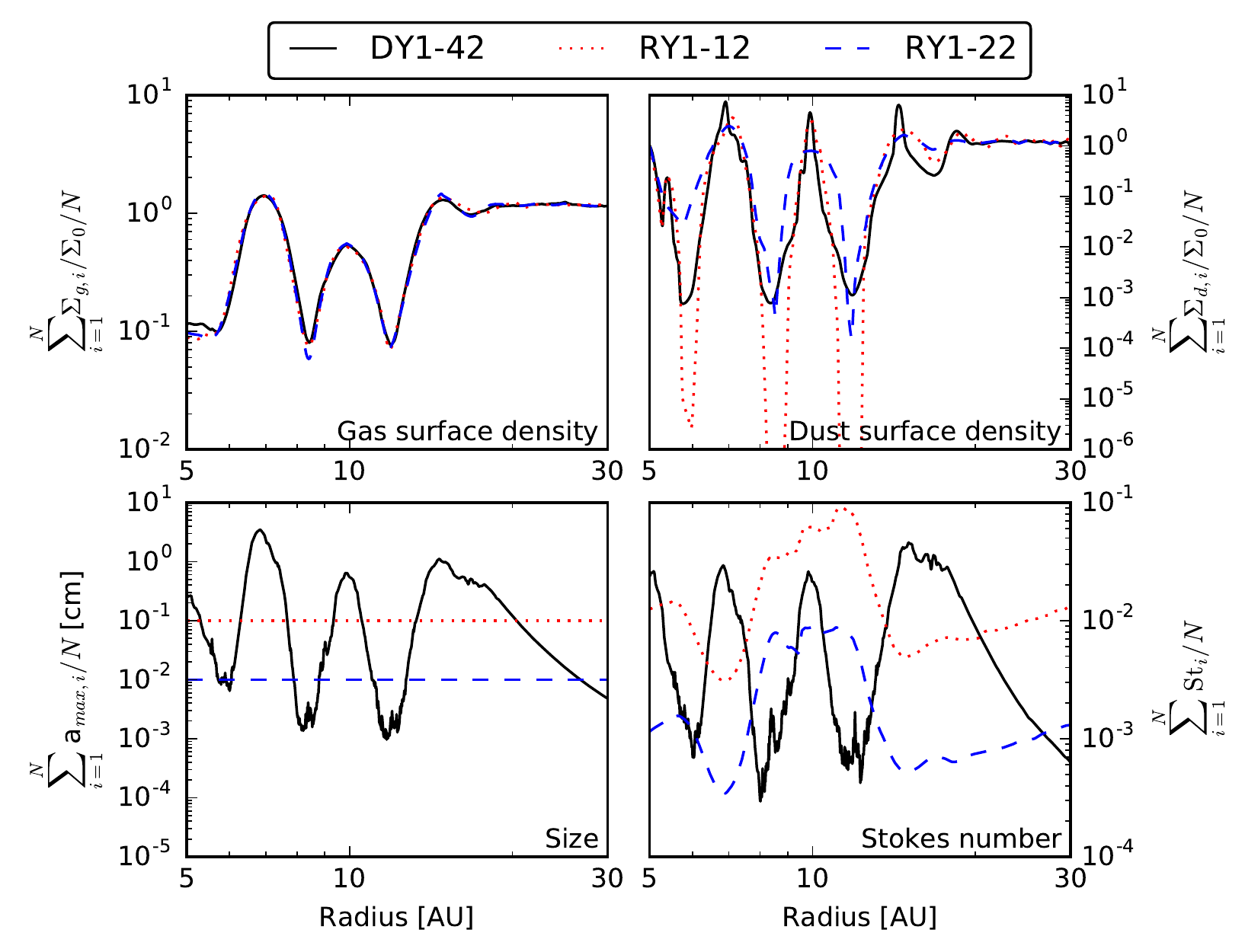}
\caption{Comparison of intermediate $10^{-1}$ cm (red dotted line) and $10^{-2}$ cm (dashed blue line) fixed particle sizes with the dust evolution simulation DY1-42 (starting size $10^{-4}$ cm, solid black line) after 400 orbits. The upper panels show the azimuthally averaged gas and dust profiles while the bottom panels show the azimuthally averaged dust size and Stokes number. In contrast to the previous comparisons: all results are of the same order of magnitude.}
\label{fig:compC_planet}
\end{figure*}

Our main focus, however, is the comparison with the new sub-grid model.
In this respect the results presented so far suggest that, when a fixed size ought to be used, small sizes (e.g. less than $10^{-1} - 10^{-2}$ cm) are a preferable choice relative to large sizes (e.g. more than 1 cm) because with the former one achieves a closer match with results obtained when the dust evolution model is employed.
This arises from the fact that particles may fragment (or drift), which leads naturally to the production of smaller particles.
There is only a small fraction of the inner disk that allows particles to grow up to 3 centimeters, hence simulations with small sizes are more realistic than those using large sizes.
Nevertheless, a more careful inspection leads to the conclusion that the simulation with small fixed size particles (RY1-42) also differs appreciably from the simulation with dust size evolution.
In the case of an evolving dust size, particles grow until they halt at the pressure-induced barrier, at which point the depletion of the inner part of the disk starts. 
Instead, in the corresponding fixed size simulation (RY1-42) this does not happen.
A possible objection to such inference could be that this particular comparison is made between
runs employing large particles (with an initial size of 3 cm), which is unlikely to be a realistic starting point. 
To address the issue further the same comparison was done with a 3 cm fixed dust size simulation RY32, a $10^{-4}$ cm fixed dust size simulation RY1-42 and a variable dust size simulation starting form $10^{-4}$ cm dust sized particles DY1-42, see Fig. \ref{fig:compB_planet}. 
In this case dust particles start growing from a realistic size and do not directly start with fragmentation.
This however, seems to favor the choice of a small grain size even more, most likely because of the same starting size of the particles.\\ 
In any case, this result should be regarded with caution. 
Indeed, we compared the three simulations only after a relatively short timescale, 400 orbits, which corresponds to only a small fraction of the typical disk evolution timescale, which is of order of Myrs.
Therefore we expect that particles would grow further after 400 orbits, thus deviating progressively more from the simulation adopting small fixed size grains. 
In order to investigate this further we employ two additional simulations, RY1-12 and RY1-22, and compare again with the DY1-42 simulation. 
These simulations do not include dust evolution but employ particles with an intermediate size ($10^{-2}$ and $10^{-1}$ cm, respectively), see Fig. \ref{fig:compC_planet}. 
Clearly these two new simulations match the dust evolution run even better, simply because at the time chosen for the comparison growth has proceeded to a stage such that the maximum dust size is closer to the size of grains in such two simulations. 
While this suggests that the match with a given fixed size simulation will depend on the time at which the comparison is carried out, it is also arguable that, in fixed size simulations, the choice of an intermediate dust size will in general yield a better match to the results of dust evolution
simulations.

\subsection{Comparison with ATHENA simulations}\label{s:ATHENA}
\begin{figure}
\plotone{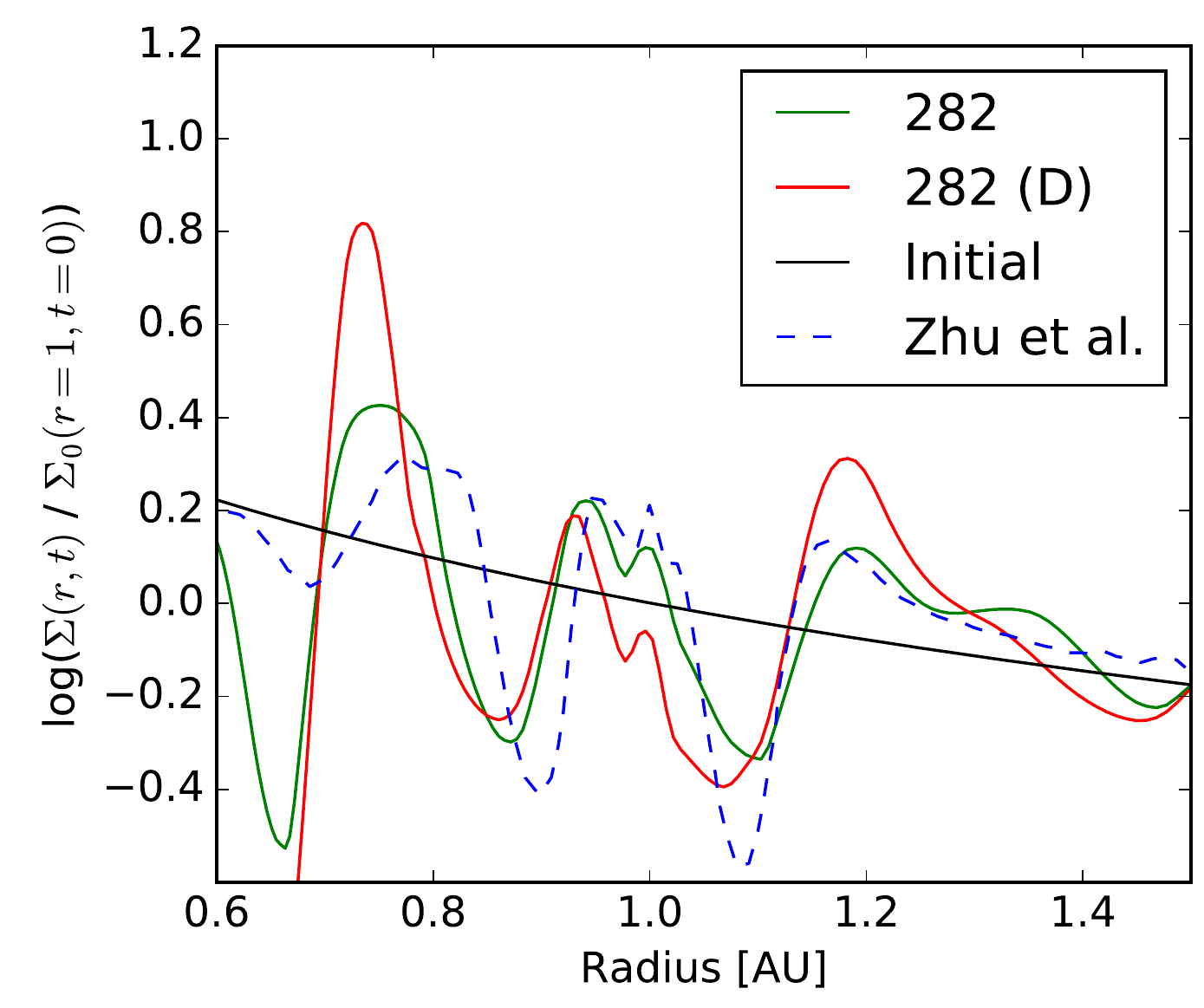}
 \caption{Reproduced dust surface density of the 'M02D2' ZH simulation after 200 orbits. A planet, thermal mass equal to 0.2 ($M_T = c^3_s /(G \Omega_K )$), was placed at R = 1. The range of the disk was 0.5 to 3 code units and an initial dust size of 2 cm was applied. The dashed blue line shows the results of ZH, the green line is the result of the fixed sized RoSSBi simulation with 282 logarithmically space grid points and the black line shows the initial conditions. The red line shows the result of the variable dust size simulation.}
  \label{fig:zhu}
\end{figure}

In this section we compare the outcome of our dust evolution model against previous results obtained
with the ATHENA code (\citealt{Stone2008}) using fixed dust size, presented by \cite{Zhu2014} (hereafter ZH). 
We choose to compare to this
particular work because the ATHENA code also uses a Godunov-type finite volume hydro method in polar
coordinates as RoSSBi.
Dust, however, is treated with Lagrangian particles advected
through the grid rather than with a second fluid as in our case. Another important difference 
is that there is no back-reaction
in the ZH simulations. We first use the RoSSBi code with a fixed dust size to reproduce 
the ZH results, and afterwards we rerun this with the dust evolution model on this particular simulation setup.
We choose to reproduce the results of the simulation 'M02D2' in the ZH paper. Hence we construct
a nearly identical disk setup with an embedded planet having a mass of 8 earth masses (using the same
thermal mass definition to initialize it as ZH, see caption of Fig. \ref{fig:zhu}), and evolve it using
an isothermal equation of state (see runs ATH and D-ATH in Tab. 1 and Tab. 2).
While the goal is to have
the same resolution as ZH in our simulation, we use a logarithmically spaced grid in the radial coordinate while ZH 
uses linearly spaced grid. Therefore, we choose the radial resolution in order to account for this difference, constructing
a grid with resolution 282 x 1024 (see Tab. 1 and Tab. 2). 
The outcome of our fixed size simulation aimed to reproduce the ZH results can be seen in Fig. \ref{fig:zhu}.
The dust density profiles are similar in shape and magnitude in comparison to the original results. 
We argue that the agreement between our results and the ZH results is sufficient to investigate the impact 
of our dust evolution scheme. Residual differences are likely caused by the different grid spacing which prevents
from comparing at truly identical resolution (we tested indeed that changing grid resolution has an effect on both
the gas and dust surface density profiles, especially near the planet).
The impact of the dust evolution method is displayed in Fig. \ref{fig:zhu} as the model 282 (D) after 200 orbits. 
It shows a significant deviation from the fixed sized simulations, in line with our findings in Section \ref{s:noplanet}, \ref{s:masterplanet} and \ref{s:mastercomparison}.
Therefore, the comparison with ZH reinforces our general inference that including dust evolution has a major effect
on the dust distribution in the disk changing the spatial distribution and amplitude of overdense regions.

\section{Discussion}

\subsection{Limitations}\label{s:limitations} 
Our numerical method is limited by several factors, which we briefly discuss here for clarity.

The original dust coagulation method was developed in the framework of an one dimensional model, whereas in this paper we use the method for two dimensional modeling. 
Therefore the coefficients of the semi analytical formulas (see Eq. \ref{eq:afrag} and Eq. \ref{eq:firstlimit}) might potentially change due to this transition, but at the moment there is no possibility to run a two dimensional hydrodynamical simulation with the full dust coagulation calculation, using a grid code, that we could use to calibrate the method. 
Hence, we adhere to the value of the coefficients proposed by \cite{Birnstiel2012}.

Another simplification that we already pointed out in Section \ref{s:Dustevo}, is that only one representative size of the dust fluid is used 
in each cell and not two sizes as proposed in the original algorithm \cite{Birnstiel2012}. 
This could change the resulting dust surface density up to a few percent.
Nonetheless, this should not impact our conclusions, which are based on the relative difference between including and not including
dust evolution.

The main challenge of our dust size treatment is that the dust size is not advected. 
The advection velocity depends on the local size and thus the density evolution is impacted by dust growth and fragmentation, but we calculate the representative size in each grid cell and in each time-step based on the local conditions at a given time only. 
The only exception is the initial growth stage, when particles grow gradually before they reach the fragmentation or drift barrier, and then their size depends on the evolution time. 
After this is completed, the particle size in a given cell changes basing on local conditions such as the dust-to-gas ratio and sound speed (see Eq. \ref{eq:afrag} and Eq. \ref{eq:firstlimit}), but it does not depend on the size that existed in the cell in a previous timestep or on the sizes obtained in the neighboring cells. 
This is equivalent to assuming that dust coagulation does always have enough time to produce an equilibrium size distribution. 
This is true in the case of a homogeneous disk, but may not be true when fast dynamical changes occur in the disk, such as the ones introduced by the planet. 
Using this model might lead to a discontinuous growth in particle size after each time step and hence affect the calculated surface densities.
This is an obvious drawback that we aim to fix in our future work.

Additionally our current code does not use any kind of sink particles and consequently the particles are trapped behind the planet but not accreted.
Introducing an accreting planet could potentially cause a decrease of the magnitude of overdensity formed behind the planet, but on the other hand the increasing planetary mass would strengthen the pressure bump that is causing the overdensity.

The limitations we mentioned above contribute to the uncertainty on the dust density evolution we obtain in our models.
However, we argue that even our approximate dust growth prescription yields a significant improvement over
simulations with fixed dust size because its shows that the difference is both quantitative and qualitative, which potentially has
paramount implication on the interpretation of observations.
A similar result has been found for an SPH based code, see \cite{Gonzalez2012, Gonzalez2015, Gonzalez2015i}.

By construction, our simulations are inviscid, and therefore no large scale turbulence is explicitly modeled. 
However, we assume that there is some level of sub-grid turbulence that triggers impact speeds between dust particles which limit their growth (see Eq.~\ref{eq:afrag}). 

Since RoSSBi solves the hydrodynamical equations using the Godunov method it also has intrinsically a very low numerical viscosity. 
The gas is in fact evolved in a very similar way as with ATHENA, concerning the un-split advection and second order time accuracy. 
However, the RoSSBi code uses an exact Riemann solver rather than HLLC-type (\citealt{Stone2008}), and a well-balanced scheme. 
In most published work codes are either inherently more viscous, such as FARGO (\citealt{Masset2000}) which is a finite difference scheme, or explicitly apply some form of viscosity to capture shocks, such as again in FARGO but also in ZEUS (\citealt{Stone1992, Stone1992a, Stone1992b}, \citealt{Clarke1996}, \citealt{Hayes2006}), or in the SPH simulations of \cite{Gonzalez2015, Gonzalez2017}. Viscosity would smooth out gaps and other features triggered by the presence of an embedded planets, generally reducing the sharpness of pressure bumps, therefore affecting the response of dust particles. This will have to be investigated in the future but, once again, here we are focusing not on absolute effects but more on the relative differences between having and not having dust evolution accounted for.

On the other hand, our method includes the back-reaction from dust to gas, which some of the other models neglect. 
This effect becomes particularly important when the dust-to-gas ratio becomes high, as the gas disk structure is modified (\citealt{Gonzalez2017}, \citealt{Kanagawa2017}). 
The back-reaction should also lead to the well known phenomenon of the streaming instability, which leads to a spontaneous clumping and formation of dusty filaments, which can subsequently become gravitationally unstable and collapse to planetesimals (\citealt{Johansen2007}, \citealt{Kowalik2013}, \citealt{Simon2016}). 
The models we present in this paper do not have enough resolution to resolve the streaming instability.
Nevertheless even a high resolution two dimensional simulation might not capture such a phenomenon, since the streaming instability is mainly an effect of the xz-plane.
However, the numerical method we developed may in fact help us in future work to revisit the streaming instability under more realistic
conditions, as all of the previous models relied on treating dust with fixed size, despite it is well known
that the streaming instability is sensitive to dust size (\citealt{Youdin2007}, \citealt{Bai2010}). Likewise, alternative
models in which overdense dust clumps are produced by vortex-drag instabilities (\citealt{CrnkovicRubsamen2015}, \citealt{Surville2016, Surville2018}), 
which would later undergo streaming instability or collapse due to their own self-gravity, will be affected
by dust evolution, although the general character (but not the timescale) of such vortex-induced instabilities
appears to be independent on dust size (\citealt{Surville2018}).

\subsection{Implications of our results}
The main purpose of this work was to compare the dust surface density evolution with dust coagulation scheme to the fixed size simulations. 
Because particle growth up to centimeter sizes is reasonably well understood and confirmed by laboratory experiments (e.g. \citealt{Guettler2010}), most of the previous simulations assumed a large grain size. 
As logical as this may sound, it leads however to a completely different evolution of the disk. 
The main problem encountered within this method is the fact that a protoplanetary disk tends to contain the largest particles in the inner part, thus the drift timescales for particles in the outer disk are slower than in simulations with a fixed size. 
Consequently, while in simulations with fixed dust size a pile-up in the inner part of the disk is not surprising, 
simulations with dust growth do not exhibit such an extremely overdense region. 
Including a planet in the simulation makes the difference even more pronounced. 
Our simple dust evolution model allows us to simulate multiple particle sizes within the dust fluid, hence also allowing small particles, which can then migrate through the planetary induced pressure bump.
In contrast to that in the fixed sized simulation, particles are held at the pressure bump in the disk. 
On the other hand, simulating a disk with small fixed sized dust particles also leads to problems. 
In this case the planet cannot stop particles from migrating; as a consequence this approach yields different results than the dust growth simulation.

As already mentioned before, it is not sufficient to choose the maximum or minimum fixed size of dust particles to reproduce the evolution of a protoplanetary disk. 
According to our results, dust growth is important and cannot be neglected. 
Since the state-of-the-art grid code simulations do not allow particle growth, it is reasonable to choose an intermediate dust particle size, which yields outcomes that are closer to the evolution that includes dust growth. 
The results ares shown in Fig. \ref{fig:compB_planet}, which compares the DY1-42 simulation, including dust growth prescription, to the fixed particle size simulations with $10^{-1}$ and $10^{-2}$ cm particles.
In these last comparison the results of the fixed dust size simulations resemble more closely those of the dust evolution simulations 
relative to the results of the RY32 
and R1-42 fixed-size simulations.
Clearly, the choice of a fixed size in order to reproduce the results of the dust evolution algorithm will depend on the duration of the
simulation. 
If a simulation starts with small growing particles and lasts only for a few thousand years, a fixed particle size that is only slightly higher than the starting size may be chosen. 
A very long simulation, may require slightly larger particles, since the sub-grid method would allow particles to grow to larger sizes as the simulation continues.
Also the inclusion of a planet affects the choice of the particle size. 
If a large planet is chosen then particles should be smaller than those selected in case of a small planet. 
This is justified by the fact that a large planet will induce a larger pressure bump which will abruptly stop larger particles. 
On the contrary, small particles can pass through the induced pressure bump for a longer time.

A major trend of our results is that, in general, when our dust evolution model is included, sharp features in the dust component
arising in disks with embedded massive planets tend to weaken significantly (see also \citealt{Zhu2012}). Such features, such as dust
rings produced by dust pile-ups near gap edges, are a recurrent feature of one-dimensional simulations that assume an axisymmetric
background (e.g. \citealt{Pinilla2012}). We argue that these previous results have largely exaggerated the strength of dust rings
and other features resulting from the perturbation of a massive planet and the formation of a gap. As a result, we caution
about using observed dust rings in disks such as HL Tau to infer the presence of a planet as the outcome is largely dependent
on the role of dust evolution (\citealt{Flock2015}, \citealt{Ragusa2017}). Here we have just begun to address this important issue with a simple-minded sub-grid model.
The implications are of paramount importance for interpretation of upcoming surveys of T-Tauri and debris disks carried 
out in various wavelengths with instruments such as ALMA and SPHERE (eg \citealt{Gonzalez2012}). As our simulations also show
a distribution of dust sizes across the disk even after 400 orbits, synthetic observations will have to be carried
in multiple wavelengths with accurate Monte Carlo radiative transfer codes in post-processing in order to provide 
a useful testbed to interpret the upcoming observations, following on the footsteps of analogous work carried out using
the RADMC3D tool for ALMA mocks of dust continuum emission from 3D spiral density waves and clumps in self-gravitating disks 
(\citealt{Dipierro2015b}, \citealt{Mayer2016}, \citealt{Meru2017}) 
as well as from 3D circumplanetary disks (\citealt{Szulagyi2018}). Indeed, while it is clear that with our dust evolution model sharp features
and transitions are smoothed out, it is also expected that this and other effects depend on the specific dust size range considered
and thus will be wavelength-dependent.

\section{Conclusions}
In this paper, we investigated the role of dust coagulation for hydrodynamical evolution of protoplanetary disk. 
Due to the complexity and computational expense of this problem, the state-of-the-art hydrodynamical grid code models would typically assume either a fixed size or fixed Stokes number for all dust particles. 
This approach is not realistic, as there are various processes determining the dust size distribution, which depend heavily on the local conditions within the disk. 
We applied a simplified treatment of dust growth and fragmentation based on an algorithm proposed by \cite{Birnstiel2012} and compared its results to the ones returned by the fixed-sized approach. 
The main conclusions of this paper may be summarized as follows:
\begin{itemize}
\item{Including the dust growth significantly changes the dust surface density evolution in protoplanetary disk. 
Assuming fixed size large particles leads to formation of exaggerated pile-ups either in the inner part 
of the disk or in a pressure bump, which are not observed if the particle sizes are more realistic. Resulting dust features
triggered by the presence of a massive planet are thus weakened, with important implications on the interpretation of recent
and upcoming high-resolution disk observations }
\item{In the more realistic models, dust growth proceeds inside-out. The large grains are first formed in the inner part of the disk, and 
their growth in the outer parts takes significantly longer.}
\item{If a fixed dust size needs to be used to limit complexity, or for computational cost considerations, our findings clearly suggest
that the correct choice consider intermediate dust sizes ($10^{-2}-10^{-1}$~cm) rather than the 
largest dust size that could be obtained by coagulation.
The specific choice, though, will depend on the evolutionary stage of the disk, as the
longer the timescale the larger is the intermediate dust size that better approximates the results of our dust evolution simulations.}
\end{itemize}
Finally, in Section \ref{s:limitations} we described the main limitations of this method.
Among the latter, the most important one is the lack of
advection of individual dust sizes.
The dust size is instead recomputed at every cell location and at every timestep based on our
current conditions. We will need to explore possible deviations from the correct dust evolution in the future by comparing with an approach in which
Lagrangian particles are introduced to represent the dust fluid, as in the ATHENA code, but dust evolution is also accounted for.
The other major avenue of development for the future will be the design of multi-wavelength
ALMA mocks of our simulations with dust evolution by means of the RADMC3D radiative transfer tool (see \citealt{Dullemond2012}).
\section*{Acknowledgments}
We want to thank the anonymous referee for the useful comments that helped us to further improve this paper. We also want thank C. P. Dullemond, T. Birnstiel,  J-F. Gonzalez, T. Henning, Z. Zhu and X. Bai for useful discussions.
JD and CS also acknowledge the support of the PlanetS National Center of Competence in Research of the Swiss National Science Foundation.

\bibliography{paper}

\end{document}